\documentclass[preprint,12pt]{elsarticle}
\usepackage{lettrine}
\usepackage{graphicx}
\usepackage{subcaption}
\usepackage{tabularx}
\usepackage{booktabs}
\usepackage{color}
\usepackage{multirow}

\usepackage{amssymb}
\usepackage{amsmath}

\usepackage{algorithmicx}
\usepackage[ruled]{algorithm}
\usepackage{algpseudocode}

\usepackage[colorlinks=true, allcolors=blue]{hyperref}

\journal{Sustainable Energy, Grids and Networks}

\begin{document}

\begin{frontmatter}



\title{Coherent Load Profile Synthesis with Conditional Diffusion for LV Distribution Network Scenario Generation}

\author[1]{Alistair~Brash \corref{cor1}}

\author[2]{Junyi~Lu}
\author[2]{Bruce~Stephen}
\author[2]{Blair~Brown}
\author[2]{Robert~Atkinson}
\author[2]{Craig~Michie}
\affiliation[2]{organization={Department of Electronic and Electrical Engineering, University of Strathclyde}, 
            city={Glasgow},
            postcode={G1 1XW}, 
            country={Scotland}}

\author[3]{Fraser~MacIntyre}
\affiliation[3]{organization={Scottish and Southern Energy Networks}, 
            city={Perth},
            postcode={PH1 3AF}, 
            country={Scotland}}
\author[2]{Christos~Tachtatzis}

\cortext[cor1]{Corresponding author at: National Manufacturing Institute Scotland, Renfrew,  PA3 2EF, Scotland. Email: alistair.brash@strath.ac.uk}
\begin{abstract}
Limited visibility of distribution network power flows at the low voltage level presents challenges to both distribution network operators from a planning perspective and distribution system operators from a congestion management perspective. More representative loads are required to support meaningful analysis of LV substations; otherwise, such analysis risks misinforming future decisions. Traditional load profiling relies on typical profiles, oversimplifying substation-level complexity. Generative models have attempted to address this through synthesising representative loads from historical exemplars; however, while these approaches can approximate load shapes to a convincing degree of fidelity, analysis of the co-behaviour between substations is limited, which ultimately impacts higher voltage level network operation. This limitation will become even more pronounced with the increasing integration of low-carbon technologies, as estimates of base loads fail to capture load diversity. To address this gap, Conditional Diffusion models for synthesising daily active and reactive power profiles at the low voltage distribution substation level are proposed. The evaluation of fidelity is demonstrated through conventional metrics capturing temporal and statistical realism, as well as power flow modelling. Multiple models are proposed to handle varying levels of data availability, ranging from unconditional synthesis to an informed generation driven by metadata and daily statistics. The results show synthesised load profiles are plausible both independently and as a cohort in a wider power systems context. The Conditional Diffusion model is benchmarked against both naive and state-of-the-art models to demonstrate its effectiveness in producing realistic scenarios on which to base sub-regional power distribution network planning and operations.
\end{abstract}



\begin{keyword}

Load Modelling, Power Systems Modelling, Neural Network Applications, Generative Modelling.

\end{keyword}

\end{frontmatter}




\section{Introduction}

Establishing the extent to which low-voltage (LV) distribution feeders are challenged is a key obstacle to achieving decarbonisation of energy end use in the form of heat and transportation. In LV networks, which in Great Britain (GB) is 415V for domestic and light commercial customers~\cite{HABEN2021117798}, visibility has become increasingly important to understand loading characteristics to prevent potential voltage and thermal excursions in the short term, and to inform network planning~\cite{Chen2014} in the long term. Adopting a monitoring regime similar to that used at the transmission level i.e., recording and storing historical values, is prohibitive owing to the number, size and heterogeneity of the LV networks - tens of thousands of 11~kV substations in a typical GB DNO licence area~\cite{Rory2021}, and in an urban conurbation such as London, there are 19,583 LV substations~\cite{ukpn2024}. Even if costs for sensing and communications infrastructure for each substation or network bus installation are fixed, consideration must be paid to the OPEX costs, such as data transmission and storage costs.

Scenario generation is one viable route to forestall potential problems from the introduction of new technology or general load growth. However, a baseline performance measure is difficult to obtain due to the lack of LV observability. At the distribution level, the primary concern is the extremes: headroom for thermal exceedances and footroom for voltage violations~\cite{10328319}. Typical Load Profiles (TLP) exist~\cite{elexon_load_profiling}, but these are generally seasonal, not weather sensitive and lack the diversity that would enable a cohort of premises to be aggregated together into a plausible LV substation load. In defining plausibility, the peak demand sharpness (and under embedded PV scenarios, trough depth) should reflect the range of time use across neighbourhood premises: too sharp would result from near identical behaviours and hence an overestimate of potential threats; too blunt assumes greater diversity than is likely, and therefore a potential overestimate of headroom. Generative modelling from a limited set of historical exemplars offers a potential solution, but the varied nature of LV substation loading introduces its own set of complexities and hurdles. The statistical distribution of LV substation load does not follow any common probability density function~\cite{Singh2009}, making sampling approaches unrepresentative unless carefully chosen for particular locations. Markov Chain approaches also struggle with capturing aggregated time use resulting from routine behaviours~\cite{kleinebrahm2021usingattentionmodellongterm}. Understanding the present loading characteristics is essential as this condition could then further worsen with the extensive installation of distributed generators as well as low-carbon heat and transport~\cite{PengUnbalanced}. This will cause equipment to operate under atypical conditions that impact the wider network~\cite{Ferreira2013}, potentially up to the transmission level. For a Distribution Network Operator (DNO), this insight is necessary for long-term planning of reinforcements that will facilitate Low Carbon Technology (LCT) adoption and generally increase resilience over time periods of months, years and decades. For a Distribution System Operator (DSO), the same insights will yield short-term generation dispatch and load curtailment potential within the constraints of the physical network infrastructure, with both active and reactive turn-up and turn-down capabilities articulated accurately via a flexibility envelope ~\cite{8291006}. This, in turn, will articulate substation flexibilities that can provide balancing at the transmission interface, which, combined, represent regional-scale scenarios from fundamental end-use insights - unlocking this knowledge of aggregating demand would enable better system planning as well as policy design over a longer period.



LV distribution network loads are much closer to individual premises end use, meaning that lower levels of aggregation will reflect highly variable behaviour routines which may result from social norms, variation in appliance ownership and specification, and also reactions to localised weather conditions. Previously, LV substation data was modelled using conventional methods such as aggregated average profiling~\cite{elexon_load_profiling} with a diversity correction. While this approach was historically sufficient, the increasing penetration of LCTs now necessitates more advanced and accurate methods. To approach this complex and non-stationary behaviour, it is proposed here that a generative AI model, specifically a diffusion-based model, is applied. This is capable of synthesising both Active and Reactive power data for LV substation load data based on a set of observed influencing factors, namely, weather and calendar variables. Diffusion is suitable for application on load data due to its ability to learn complex underlying temporal and statistical correlations present in load profile data. Furthermore, a conditional diffusion model can make use of substation metadata along with weather and calendar variables, enabling higher-quality load profile scenario generation. Diffusion is also the state-of-the-art method for the generation of time series data.

Multiple diffusion models are proposed throughout this study, each with their own set of conditional variables. This design choice is motivated by two main factors: variability in data availability across substations and the expected gains in generative fidelity offered by conditioning. Firstly, in reality, available data will vary across substations, ranging from limited monitoring to basic metadata or, in some cases, no information at all. This study aims to provide multiple models capable of handling all these scenarios, both for practical deployment reasons and to demonstrate the expected improvements in generative fidelity that arise from conditioning on additional information. The proposed model can capture and replicate realistic substation-specific behaviour by conditioning on lightweight information such as basic metadata or daily summary statistics, without the significant overhead of mass data collection and storage. These models will also be capable of supporting different use cases, such as scenario generation under limited data availability or unconditional and unrestricted data generation for complete hypothetical network generation, which will be outlined later in this study.

In this study, coherence is not defined as the joint synthesis of an entire LV network, nor as explicit coordination between substations during generation. Instead, it is treated as an emergent property arising from the aggregation of independently generated LV substation load profiles. The proposed diffusion models synthesise individual substation active and reactive power profiles with realistic temporal structure and diversity learned from historical data, without enforcing synchronisation.

Coherence is therefore evaluated \emph{a posteriori} via power flow modelling, by propagating the independently synthesised substation loads through a sample distribution network and examining the resulting voltage magnitudes and phase angles. This analysis serves as a test of whether the generated loads collectively produce physically plausible and operationally meaningful network behaviour. Synchronisation would exaggerate voltage and thermal stress, while excessive smoothing would mask potential constraint violations. The objective is thus to generate substation-level loads whose diversity is sufficient to yield realistic network responses when analysed using standard power system tools.

\subsection{Summary of Contributions}
This section will outline the key contributions made in this study. These are as follows:

\begin{enumerate}
    \item \textbf{Application of a deep generative diffusion-based model to LV substation load modelling} - To the best of the authors’ knowledge, this study represents the first application of diffusion-based generative modelling to load synthesis at the LV substation level, addressing a scale and use case not considered in prior diffusion-based load studies, which have primarily focused on other parts of the power system.
    \item \textbf{Synthesis of reactive power at the LV substation level} - This study is the first to extend LV substation load synthesis to reactive power. Modelling reactive power is becoming increasingly important as it captures the aggregated inductive and capacitive characteristics of devices connected at the LV level and is therefore a critical quantity for realistic LV load modelling; as emerging technologies are increasingly connected to the grid (e.g., heat pumps, EVs, and other low‑carbon technologies), reactive power is expected to become more variable at the LV level, such that the previously common assumption of a constant power factor no longer holds.

    \item \textbf{Power system analysis of synthesised LV load profiles} - Unlike many existing studies, realism is not assessed solely through statistical similarity of the individual load profiles. Instead, the proposed approach explicitly examines how the co-occurrence of load behaviours across cohorts of independently synthesised LV substations propagates through the network and affects voltage magnitude and phase angle behaviour at higher voltage levels using a sample network. This power-system based evaluation tests coherence and co-behaviour in the power systems context, which results in a more realistic operational consequence on which to base a planning or investment decision.
\end{enumerate}

This paper is structured as follows: Section~\ref{ssec:litreview} reviews the related work on load synthesis tasks. Section~\ref{sec:methodology} proposes a generative approach to LV substation load data synthesis based on contextual conditions. Section~\ref{sec:results} evaluates the performance of the model through multiple synthesis metrics and visualisations. Section~\ref{sec:casestudy} conducts a Load Flow analysis to evaluate wider power system coherence through the substation hierarchy. Section~\ref{sec:conclusion} presents the conclusions and future work of the study.




\section{Related Work}\label{ssec:litreview}

Approaching an energy system in transition, it is important to be able to identify the capabilities and limitations of infrastructure already in place, with its present-day utilisation. Understanding scenarios in the present allows the ability to accommodate new technologies to be anticipated. In the past, methods of average profiling have been used to model the load behaviour of electricity end use, such as system operator typical load profiles (e.g ~\cite{elexon_load_profiling}) or clustering approaches~\cite{5944675, 8973532}. These methods leverage highly generalised `typical' customer behaviours to provide a series of averaged profiles which incorporate factors such as building type, seasonal trends, days of the week, etc. However, these methods remain static and lack diversity when modelling the collective load on the substation in which a cohort of them is connected. Realistic modelling requires an approach that captures the diversity inherent in behaviour profiles as opposed to those generated via coarse-grained averaging. 

One approach is Gaussian Mixture Models (GMMs)~\cite{ZHANG20201221, LI2018331, 7322269}, that involves fitting a mixture of Gaussian distributions to model load profiles based on a number of components (the number of Gaussian distributions fitted across the dataset). The optimal number of components can be calculated using the model complexity metric, Bayesian Information Criterion (BIC), to provide the best fit for the given dataset. Although GMMs improve upon the traditional average load profile approaches by capturing more realistic load profiles, they lack the temporal and statistical accuracy essential for LV substation load monitoring.

Deep learning approaches to load synthesis have been proposed for smart meter data through the use of Generative Adversarial Network (GAN) models and Convolutional Variational Autoencoders (CVAEs)~\cite{chai2024faradaysyntheticsmartmeter, hu_multiload-gan_2024, yilmaz_synthetic_2022}; there are variants whereby these deep learning models have been combined with clustering methods~\cite{9139305, WANG2020110299}. More recently, applications of Diffusion models have been applied to power systems data; however, not in the context of load modelling at the LV substation level. Source–load scenarios for renewable generation are synthesised at a power systems level using diffusion models~\cite{ZHAO2025124555, DONG2025126446}. Diffusion models have, however, been applied to load modelling at the smart-meter level of the network~\cite{wang2024customizedloadprofilessynthesis}, and net-load synthesis has also been applied to customer data~\cite{zhang2024generatingsyntheticnetload}. 

Despite the similarities between smart meters and LV networks, these approaches are not directly applicable to LV networks due to the broader range of factors that must be considered, such as weather and LCT penetrations, for example. These considerations result in load profiles at different levels of the network having different underlying statistical distributions. Furthermore, although the studies have shown excellent ability to synthesise individual smart meter load profiles~\cite{chai2024faradaysyntheticsmartmeter, hu_multiload-gan_2024, yilmaz_synthetic_2022, 9139305, WANG2020110299}, they have no consideration of diversity across cohorts of premise meters. Consequently, the quality of the aggregated profile at the next voltage level up may lack representativeness. Additionally, these models are often developed to serve different objectives such as increasing data quantity due to availability issues stemming from privacy~\cite{chai2024faradaysyntheticsmartmeter, hu_multiload-gan_2024, yilmaz_synthetic_2022, 9139305, WANG2020110299}. The purpose of the proposed study is to synthesise load profiles based on individual characteristics rather than just typical scenarios. The addition of extra conditions (e.g., daily min, mean and max power, and substation-specific metadata) alongside the typical conditions observed in the literature (weather and calendar variables) allows the new model to fulfil this specific goal ~\cite{chai2024faradaysyntheticsmartmeter, hu_multiload-gan_2024,  WANG2020110299}. 

The proposed conditional diffusion model permits load profile synthesis at low aggregations on distribution networks with conditions such as weather, statistics and calendar variables that address the challenge posed by diversity when utilising average profiling approaches. Furthermore, the ability to synthesise reactive power enables load flow analysis, which allows evaluation of wider power system coherence.

\section{Methodology}\label{sec:methodology}

This section proposes the diffusion models developed to synthesise load profiles with an aim of providing operational data with realistic variability dependent on various factors such as weather data, calendar variables, and behavioural trends. Firstly, the concept of diffusion models and the SSSDS4 model is introduced in Section~\ref{sec:diffusionmodels}, then the developed models are presented in Section~\ref{ssec:lps}, and the dataset used is described in Section~\ref{sec:datadesc}. Finally, the evaluation metrics and visualisations used are outlined in Section~\ref{ssec:metsandvis}.

\subsection{SSSDS4 Conditional Diffusion Model}\label{sec:diffusionmodels}

Diffusion models were selected as the deep learning generative framework because their modelling assumptions align well with the statistical and temporal characteristics of LV substation load data. LV load profiles are high-dimensional, strongly time-dependent, and exhibit non-Gaussian, often multi-modal distributions driven by aggregated behavioural routines and external conditions. Diffusion models have demonstrated strong performance in time-series generation and imputation tasks across multiple domains, consistently outperforming GANs and CVAEs in capturing complex temporal structure while preserving sample diversity~\cite{alcaraz2023diffusionbased, NEURIPS2021_cfe8504b, chen2020wavegradestimatinggradientswaveform, NEURIPS2021_49ad23d1, NEURIPS2020_4c5bcfec}. These improvements were demonstrated across multiple datasets in different domains, showing the strong robustness and generalisation of the models. Diffusion models are particularly applicable to load data due to their ability to model high-dimensional data distributions while maintaining their diversity~\cite{wang2024customizedloadprofilessynthesis}. These properties are particularly important for LV load synthesis, where a realistic representation of diversity is required to avoid both excessive synchronisation and unrealistically smooth aggregation effects. Accordingly, diffusion models were chosen as a principled and suitable framework for the specific problem of LV substation load synthesis.

Conditional diffusion was adopted to exploit contextual information that is typically available in distribution network settings, such as calendar variables, weather data, and basic substation metadata, to guide the generation process. While diffusion models provide improved diversity relative to GAN-based approaches~\cite{wang2024customizedloadprofilessynthesis}, accurately representing the heterogeneity and substation-specific behaviour of LV loads remains challenging in an unconditional setting. Introducing conditioning constraints the generative process to physically and contextually plausible regions of the data space, enabling more faithful representation of substation-level diversity and improved reconstruction accuracy. The effectiveness of conditional diffusion for load-related modelling tasks has also been demonstrated at other levels of the power system ~\cite{ZHAO2025124555} outside the LV level of aggregation.

\begin{figure}
    \centering
    \includegraphics[width=0.9\linewidth]{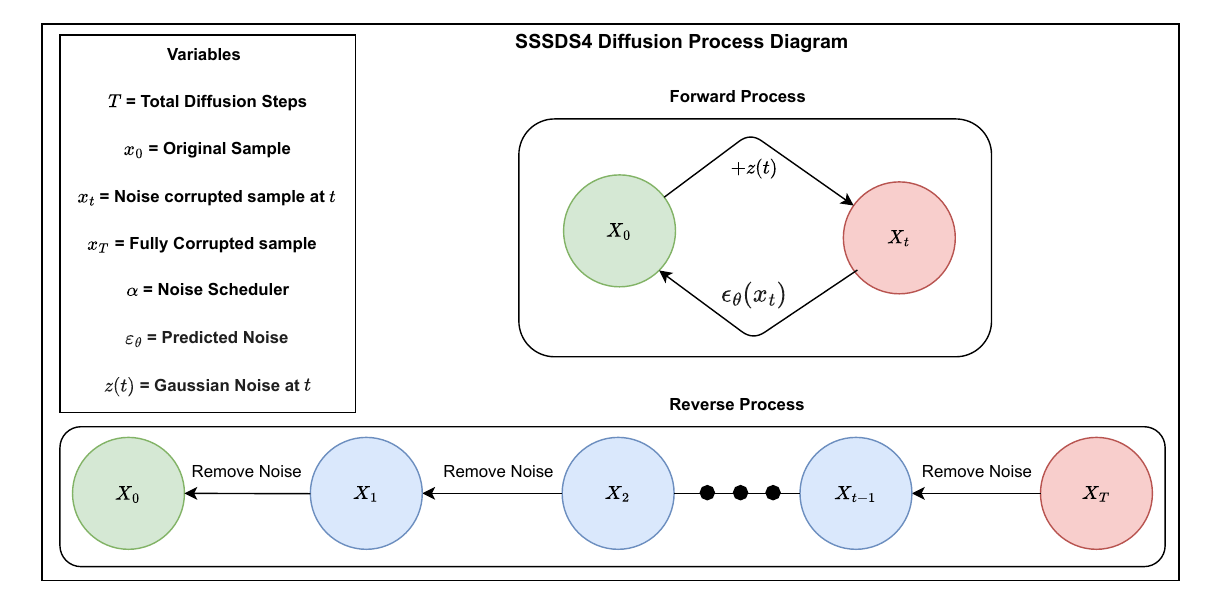}
    \caption{Diagram outlining the forward and reverse process of the diffusion model, key variable names are provided with the diagrams.}
    \label{fig:DiffusionProcess}
\end{figure}

Diffusion probabilistic models generate new independent samples that statistically resemble the distribution of the underlying data on which they are trained. They operate in a forward (training) and reverse inference process. The diffusion forward process follows a Markov chain where Gaussian noise is progressively added to a sample \(x_0\) in $T$ steps; $t = 1, \ldots, T$. The level of noise \(z(t)\) added in each step \(t\) is typically drawn with a linear noise scheduler \(\alpha\) between an upper and lower noise level \(\beta_0 \, \text{and} \, \beta_t\) resulting in a corrupted sample \(x(t)\).  During training, rather than simulating the multiple steps of the Markov chain for every training iteration, it is more efficient to corrupt the sample in a single stage for all steps \(t\) by sampling directly from cumulative Gaussian noise, which can be calculated using  \(\alpha\)~\cite{NEURIPS2020_4c5bcfec}. Figure~\ref{fig:DiffusionProcess} illustrates the forward process of the model, where for a randomly sampled diffusion step  \(t\in(0,T]\). The model corrupts the sample with \(z(t)\) and predicts \(\epsilon_\theta (x_t)\), the level of the cumulative noise in \(x_t\) which accounts for the noise added in all steps \(0\to t\). The model is trained for multiple iterations and learns to predict noise for all values of  \(t\in(0,T]\).  During inference, the reverse process recovers the sample \(x_0\) from \(x_T\), where the latter is pure Gaussian noise. The inference process also follows a Markov chain but in the reverse direction, from steps \(t = T, \ldots, 0\). The noise is iteratively removed by first predicting the cumulative noise from the current step \(x_t\) to \(x_0\) - to the beginning of the chain \( \epsilon_\theta(x_t)\). Then the model removes the portion of the noise attributed to step \(x_t\) to \(x_{t-1}\), which can be estimated using Equation~\ref{eq:noise_removal}. Finally, at each step, a variance term is added to \(x_t\) to retain stochasticity in the generation process and permit synthesis of diverse samples~\cite{tashiro2021csdi, lin2023diffusion}.

\begin{equation}\label{eq:noise_removal}
x_{t-1} = \frac{x_t - \frac{1 - \alpha_t}{\sqrt{1 - \bar{\alpha}_t}} \cdot \varepsilon_\theta(x_t, t)}{\sqrt{\alpha_t}}+\sigma_t
\end{equation}


The proposed diffusion model is built upon an implementation of the SSSDS4 model~\cite{alcaraz2023diffusionbased}. SSSDS4 provides a time series implementation of a diffusion model and was shown to achieve strong performance in several generation tasks across multiple datasets when benchmarked against other diffusion-based approaches~\cite{alcaraz2023diffusionbased}. SSSDS4 was enhanced over existing methods~\cite{tashiro2021csdi} through the incorporation of structured state space models, to allow capturing long-term dependencies in the data, such as seasonal and diurnal trends linked by weather and behavioural routine~\cite{abs-2111-00396}. These are known to be key factors in load behaviour~\cite{shi2014analysis} and were the key motivation in the selection of the SSSDS4 as the diffusion model to apply to the LV substation synthesis.

The SSSDS4 model was adapted to utilise conditional inputs to constrain the generation process and synthesise samples from specific areas of the overall distribution. Details of the masking mechanism are outlined in \ref{sec:appmask}. Furthermore, the SSSDS4 model source code contained an implementation error in the training loop relating to the handling of conditional inputs. Algorithm~\ref{Algorithm:ddpm_training_masked} outlines the modified training step in the SSSDS4 model. There is one key difference between this and the previous model. Line 7 restores the conditional values after the entire sample is corrupted with noise. Previously, this was handled differently; the conditional values were substituted into \(x_{std}\) (the noise vector). Then, when the sample is corrupted, the conditional values would not be corrupted with noise. However, the noise corruption equation outlined in line 6 only balances the data for values of \(x_0\) \(\in [0, 1]\). In many cases, this is not sufficient; a more effective method is to corrupt the entire sample with noise, then replace the values after corruption. This aligns with the behaviour during inference, where the conditional values are replaced at each step.

\vspace{-0.2cm}%
\alglanguage{pseudocode}
\begin{algorithm}[h]
\small
\caption{SSSDS4 Modified Training Step}
\label{Algorithm:ddpm_training_masked}
\begin{algorithmic}[1]
\Require{$\{T, \beta_0, \beta_1\}$ (diffusion hyperparameters), $\text{Net}$ (model), $m_{\text{imp}}$ (imputation mask), $m_{\text{mvi}}$ (missing value mask), $\mathbf{x}_0$ (ground truth signal)}{}

    \State $a_t$ \Comment Noise schedule coefficient for step $t{-}1 \to t$
    \State $\bar{\alpha}_t = \prod_{s=1}^t \alpha_s$ \Comment Cumulative product of alphas

    \State $x_{\text{std}} \sim \mathcal{N}(0, 1)^{C_{\text{size}}}$ \Comment Generate initial Gaussian noise
    \State $D_S \sim \text{RandInt}(T)^{C_{\text{size}}}$ \Comment Sample diffusion step indices

    \State $M \gets m_{\text{signal}} \odot m_{\text{condition}}$ \Comment Mask for conditional inputs

    \State $\bar{x} \gets \bar{\alpha}_t[D_S] \cdot x_0 + (1 - \bar{\alpha}_t[D_S]) \cdot x_{\text{std}}$ \Comment Corrupt Sample with noise
    \State $\bar{x} \gets (\bar{x} \odot (1 - M)) + (x_0 \odot M)$ \Comment Restore conditional values

    \State $C \gets \text{Concat}(x_0 \odot M, M)$ \Comment Conditioning input to network
    \State $y \gets \text{Net}(\bar{x}, C, D_S)$ \Comment{Predict noise present in \(\bar{x}\)}

    \State $\text{loss} \gets \left\| y \odot M - x_0 \odot M \right\|^2$ \Comment MSE loss over masked positions
    \State \textbf{update\_parameters}($\text{loss}$)

\Statex
\end{algorithmic}
\vspace{-0.4cm}%
\end{algorithm}

\begin{figure*}[!htbp]
    \centering
    \includegraphics[width=0.9\textwidth]{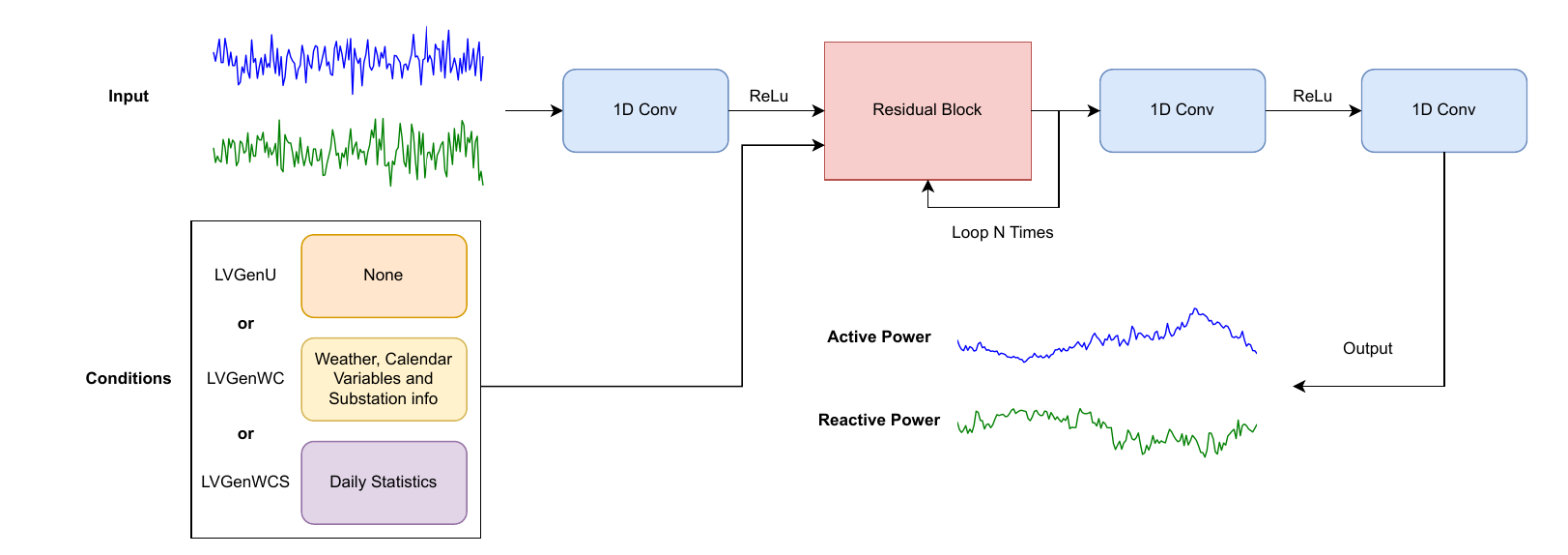}
    \caption{SSSDS4 Model implementation outlining model inputs, outputs and key layers/blocks.}
    \label{fig:sssds4}
\end{figure*}

\begin{figure}[!htbp]
    \centering
    \includegraphics[width=0.9\textwidth]{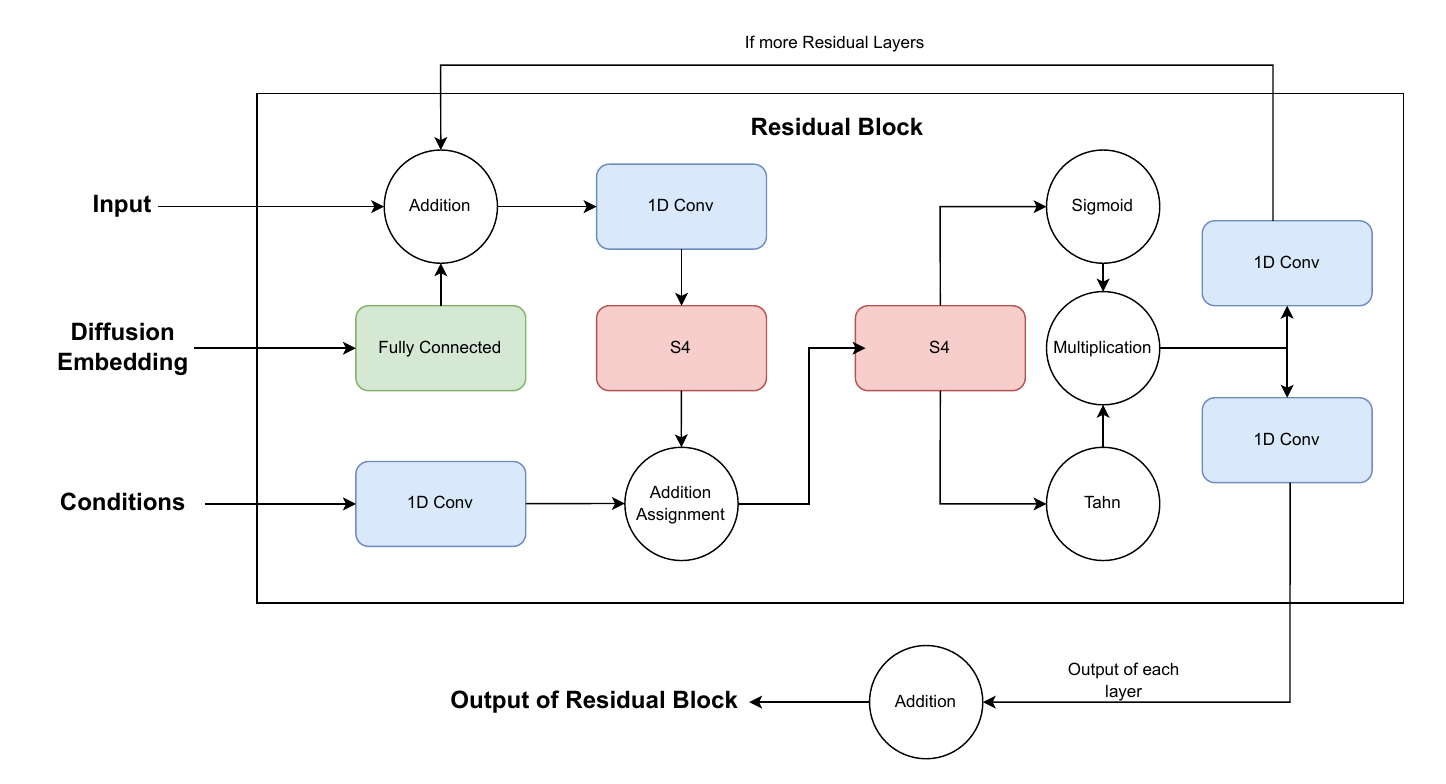}
    \caption{Diagram of the Residual Block within the SSSDS4 Model.}
    \label{fig:rb}
\end{figure}

Figure~\ref{fig:sssds4} shows a diagram for the implementation of the SSSDS4, highlighting the model inputs/outputs and key layers. The minor modification made to the model can be observed in the inputs and conditions being passed to the model. Figure~\ref{fig:rb} provides more detail of the residual block section of the diagram. The block represents the structure of a single residual layer, and multiple of these blocks are chained together. After a hyperparameter search, 36 residual blocks were used. The output of all residual blocks is then added together and fed to the subsequent 1D Convolutional layer.

\subsection{Load Profile Synthesis Models}\label{ssec:lps}

This study proposes three diffusion models:

\begin{enumerate}
    \item LVGenU - The unconditional diffusion approach provides a blind synthesis without utilising conditions. In this form, the diffusion model attempts to recreate load profiles from Gaussian noise alone.
    \item LVGenWC - This model adds conditional variables which require no daily storage; thus, more accurate daily loads could be reconstructed without the requirement of any recorded values. The features used for this model are as follows; calendar information (day of week, month etc.) which can be directly inferred from the timestamp, weather forecast (temperature, humidity, wind speed, etc.) collected from API based on the substations latitude/longitude co-ordinates, and substation information (number of customers connected) which will be stored for each substation separately.
    \item LVGenWCS - The next addition of cues involved passing the daily minimum, mean, and maximum of the active and reactive power. This allows for a significant improvement of synthesis and reconstruction, particularly in the more extreme values from substations with unique behaviours. 
\end{enumerate}

\begin{figure}[!htbp]
    \centering
    \includegraphics[width=0.7\textwidth]{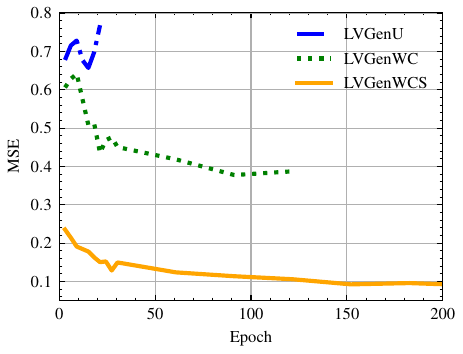}
    \caption{Test Loss for each Model through epochs.}
    \label{fig:conver}
\end{figure}

Each model was trained with the optimal neural network configuration defined in~\cite{alcaraz2023diffusionbased}, with 200 diffusion steps. Each model was trained with a maximum of 200 epochs, but was halted earlier if test loss plateaued. Figure~\ref{fig:conver} shows the convergence of each model based on the MSE vs the number of epochs; LVGenU converges relatively quickly as training terminates when no improvement in validation loss is observed for 2 consecutive checkpoints. A short patience was selected because the unconditional model rapidly reaches a performance plateau, after which further training does not yield measurable improvements and risks overfitting rather than learning additional structure. This occurs because there are no conditions for the model to learn from; LVGenWC and LVGenWCS require a significantly higher number of iterations before the models converge. Full details of the model configuration are available in \ref{sec:appconfig}.

\subsection{Data Set Description}\label{sec:datadesc}
Distribution substation monitoring data was acquired from the National Grid Electricity Distribution (NGED) OpenData~\cite{national_grid_connected_data_portal} platform. It constitutes a subset of monitored LV Networks consisting of 1,431 substations, totalling approximately 19,050 days of data (as of 6$^{th}$ of June 2024). Measurements include the active and reactive power for each phase (L1, L2, L3). Each substation was paired with associated metadata, including primary substation identifier, number of connected customers, weather variables obtained from the Meteostat API~\cite{meteomatics_weather_api}, and calendar information. Records with missing or incomplete samples were discarded. Data cleansing further removed daily profile outliers (e.g., days with unrealistically high load values exceeding 1,000 kW), incomplete days, and anomalous profiles such as constant values recorded throughout the day. If any of these criteria were met, the entire daily load profile was removed. This resulted in a cleansed dataset of 13,756 days.

The dataset was split into training and test sets at the substation level. For each substation, the total number of available daily profiles was recorded, and substations with fewer than 10 days of data were assigned to the test set, while the remaining substations were assigned to the training set. This threshold was chosen to mitigate overfitting/memorisation during training. Using a cut-off of 10 days resulted in an approximate 70/30\% train–test split (69.93\% of samples in the training set). This strategy also prevents data leakage, as all observations from a given substation are contained exclusively within a single split.

The number of connected customers was discretised into fixed-width bins (0–99, 100–199, etc.), with a final bin aggregating substations with more than 600 customers. This discretisation was adopted to prevent the model from memorising or implicitly regressing on the exact customer count, which would reduce generalisation and risk information leakage across substations with similar sizes. This upper bin was introduced to avoid sparsely populated bins at the upper end of the customer-count distribution, as fewer than 20 substations exceeded this threshold (representing approximately 0.83\% of all substations, with a maximum of approximately 800 customers).

Finally, in order to permit testing of the load flow simulation presented in Section~\ref{sec:casestudy}, a set of substations which are connected to the same Primary Substation were held back from the testing set.

\subsection{Evaluation Metrics and Visualisations}\label{ssec:metsandvis}

In this study, two metrics are used to evaluate the performance of the data synthesis models in two key areas. Mean Squared Error (MSE) is used to measure the reconstruction errors of the generated load profiles vs the actual. MSE was chosen as it is an error metric that is routinely used with machine learning applications. The Maximum Mean Discrepancy (MMD) is used to measure the quality of the synthesis. MMD is a non-parametric, kernel-based distance metric that quantifies differences between real and generated probability distributions by comparing their means in a reproducing kernel Hilbert space~\cite{10.5555/2188385.2188410}. LV substation load data are non-Gaussian and do not follow a common parametric probability density function~\cite{Singh2009}, rendering parametric divergences (e.g., KL divergence) unsuitable or unstable in this context. MMD was therefore selected as it enables robust comparison of complex, multi-modal load distributions without imposing restrictive distributional assumptions. For this reason, MMD is widely adopted in data synthesis studies involving non-Gaussian and high-dimensional data~\cite{gao_two_nodate}.

The remainder of this section will focus on MSE and MMD metrics; however, additional metrics are also reported, mainly inspired by previous research in load synthesis, to provide further insights into model performance~\cite{9534373, li2024diffchargegeneratingevcharging}. These further distance-based metrics include Wasserstein distance~\cite{doi:10.1137/1118101}, which can further evaluate modelling at the distribution tail, marginal score~\cite{li2024diffchargegeneratingevcharging}, which models the distributional alignment based on a fixed number of bins, and MiVo~\cite{9534373}, which provides a distance-based metric with additional awareness to diversity. These metrics can be used alongside the core metrics to better understand the modelling of key areas of the distribution and will be referenced where appropriate. An additional tail-aware MSE metric is proposed to quantify discrepancies between generated and real data at distributional extremes. This is computed by restricting the MSE calculation to observations exceeding the 95th percentile of the empirical distribution. This metric provides targeted insight into the modelling of extreme operating regions, which are particularly relevant in load modelling to avoid issues such as thermal exceedances and voltage collapse. This metric is referred to as the Tail-Focused MSE.

Each metric is computed on the full test set for both active and reactive power, with the average value reported as the final score. The MSE, Wasserstein distance, and Marginal Score are computed on daily aggregated distributions collated across all test-set substations, providing a global assessment of synthesis quality. In contrast, the MMD and MiVo metrics are computed on daily profiles of individual substations for the joint distribution of active and reactive power, with the reported value corresponding to the average across all substations. The MSE and Wasserstein distances are computed using standard Python implementations from scikit-learn (v0.22.2) and SciPy (v1.16.3), respectively. For MMD and Marginal Score, the methodology described in~\cite{li2024diffchargegeneratingevcharging} is adopted, while MiVo is computed following the approach outlined in~\cite{9534373}. Separate performance of active and reactive power on each metric is available in Appendix~\ref{sec:AdditionalResults}.

Data visualisations are used alongside these metrics to further evaluate the quality of the generated samples. The first kind of visualisation used is distribution plots, which contrast the implied probability distribution of the actual data and each proposed model. A zoomed-in section of the tail of the distribution is also provided to inspect the important extreme values, which in operational scenarios could result in problems such as voltage collapses. The second method of visualisation is demonstrated through decile plots, which plot the values of the load at various deciles of the implied probability distribution. The decile plots (10th–90th percentiles) are calculated across substations by extracting the corresponding decile at each timestep. This produces representative load profiles for each decile, illustrating variability and tail behaviour over the diurnal cycle. The final visualisations used are Autocorrelation Function (ACF) plots, which measure similarities in temporal dependence between real and generated datasets. The ACF is computed independently for each substation load profile by correlating the signal with lagged versions of itself. This is performed at each 10-minute interval of the daily load profile, resulting in 143 lag values per substation. The ACF values are then averaged across substations to provide an aggregate comparison of temporal dependency structure between models. Similar ACF plots indicate that datasets share a similar underlying temporal structure, meaning they exhibit comparable diurnal patterns or intra-day dependencies. Thus, the resemblance in ACF plots indicates that the generated datasets preserve similar temporal dependency patterns to those observed in the real data~\cite{box2015time, yilmaz_synthetic_2022}.

\section{Synthesis Results}\label{sec:results}

This section describes the performance of the proposed models through the use of various metrics and visualisations. Proposed model results are benchmarked against a Gaussian Mixture Model (GMM) and a Wasserstein GAN (WGAN). These benchmarks are mainly used for comparison with the unconditional diffusion model, after which conditional diffusion models are introduced to demonstrate the enhancements achieved through conditioning. In the subsequent section, the Tao Vanilla model is added as a further benchmark. This model could not be applied in this section as it must be trained on the specific target substations. Given that the test set consists of unseen substations, it is not possible to train the Tao vanilla and provide a fair comparison.

Random samples from GMMs provide a good model to compare metrics against the diffusion approach due to their generative nature and their previous use in load synthesis tasks~\cite{ZHANG20201221, LI2018331}. GMM can be fitted against the data set distribution using the optimal number of components derived from the Bayesian Information Criterion (BIC), which for this data was one component for the active power and four for the reactive power. The samples can then be generated by drawing samples from the resulting GMM at random.

A Wasserstein GAN (WGAN) is trained as a comparable deep learning approach. Information regarding the training methodology and procedure is described in~\ref{sec:app}.

\begin{table}[htbp]
    \centering
    \resizebox{\textwidth}{!}{%
        \begin{tabular}{lccccc}
            \toprule
            \textbf{Metric} & \textbf{GMM} & \textbf{WGAN} & \textbf{LVGenU} & \textbf{LVGenWC} & \textbf{LVGenWCS} \\
            \midrule
            \textbf{MSE} 
                & 0.93 ($\pm$2.2) 
                & 1.3 ($\pm$2.4) 
                & 0.68 ($\pm$1.6) 
                & 0.29 ($\pm$1.2) 
                & \textbf{0.10 ($\pm$0.4)} \\

            \textbf{MMD} 
                & 0.54 ($\pm$0.53)  
                & 0.67 ($\pm$0.44) 
                & 0.44 ($\pm$0.44) 
                & 0.13 ($\pm$0.25) 
                & \textbf{0.005 ($\pm$0.005)} \\

            \textbf{Marginal Score} 
                & 0.24 
                & 0.09 
                & 0.16 
                & 0.12 
                & \textbf{0.06} \\

            \textbf{MiVo} 
                & 0.73 
                & 0.97 
                & 0.57 
                & 0.21 
                & \textbf{0.05} \\

            \textbf{Wasserstein Distance} 
                & 0.29 
                & 0.28 
                & 0.37 
                & 0.15 
                & \textbf{0.015} \\

            \textbf{Tail-Focused MSE} 
                & 7.61 
                & 7.56 
                & 7.70 
                & 4.32 
                & \textbf{0.65} \\
            \bottomrule
        \end{tabular}%
    }
    \caption{
    Error metrics for each model. Results are calculated on the entire unseen test set using scaled values. The results represent the average of each metric's performance calculated separately on active and reactive power, respectively. Some metrics contain the  $\pm$ standard deviation from the mean on a per-substation basis. Best scores are highlighted in bold.
    }
    \label{tab:errors}
\end{table}

Table~\ref{tab:errors} reports the values of each evaluation metric for all models considered. The unconditional models are optimised according to fundamentally different objective functions, which directly explains their relative performance across metrics. LVGenU is trained to minimise MSE at the level of individual substation load profiles, prioritising accurate reconstruction and temporal coherence. In contrast, the GMM is optimised via expectation–maximisation to maximise the log-likelihood of the aggregated load distribution across all substations, while the WGAN is directly optimised to minimise the Wasserstein distance between aggregated load distributions. As a consequence, each model performs most strongly on metrics that align with its underlying optimisation objective.

\begin{figure*}[!tb]
    \centering
    \begin{subfigure}[b]{0.48\textwidth}
        \centering
        \includegraphics[width=\textwidth]{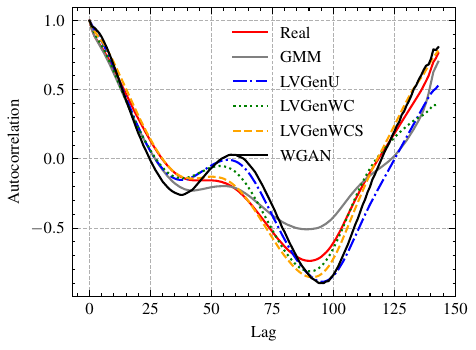}
        \caption{Active Power}
        \label{fig:acfap}
    \end{subfigure}
    \hfill
    \begin{subfigure}[b]{0.48\textwidth}
        \centering
        \includegraphics[width=\textwidth]{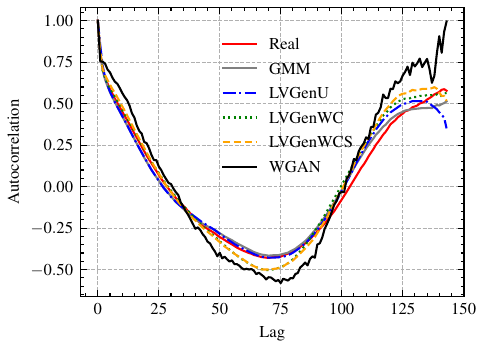}
        \caption{Reactive Power}
        \label{fig:acfrp}
    \end{subfigure}
    \caption{Comparison of ACF plots for active and reactive power data.}
    \label{fig:acf_comparison}
\end{figure*}

Consistent with this, LVGenU achieves lower values for MSE, MMD, and MiVo relative to the GMM and WGAN baselines. The reduced MSE indicates improved reconstruction fidelity at the substation level, while the lower MMD reflects closer alignment between the joint distribution of generated active and reactive power profiles and the corresponding real data. The lower MiVo score further indicates improved preservation of variability and co-dependence between active and reactive power across individual substations. Together, these metrics demonstrate that LVGenU more effectively preserves the temporal structure of individual load profiles, which is essential for maintaining realistic voltage trajectories and dependencies between active and reactive power, a trend that is also reflected in the autocorrelation function shown in Figure~\ref{fig:acf_comparison}.

At the same time, LVGenU does not outperform the other unconditional baselines on metrics that primarily assess distributional shape, including the Marginal Score, Wasserstein distance, and Tail-Focused MSE. In these cases, the GMM and WGAN can achieve lower scores, reflecting their explicit optimisation for aggregated distributional similarity. In particular, the WGAN incorporates the Wasserstein distance directly into its loss function, enabling improved global distribution capture at the expense of substation-level temporal realism and coherence. This highlights a clear trade-off between capturing global distributional shape and preserving substation-level temporal realism.

Overall, LVGenU does not capture the distribution of the aggregated load profiles but exhibits strong temporal coherence and reconstruction fidelity. These characteristics are particularly important for realistic LV load synthesis, where preserving substation-level dynamics is more critical than matching aggregated distributional shape. However, misrepresentation of extreme operating conditions may lead to an underestimation of peak loads, which in a power systems context could result in the failure to model critical scenarios such as thermal excursions. These limitations motivate the extension of LVGenU through additional conditioning in the proposed framework."

The introduction of conditions in the further models results in a significant improvement in all metrics and, in turn, mitigates the issues in the LVGenU model. LVGenWC is able to build on the performance of LVGenU, improving in all metrics. The conditions not only improved on the struggling areas of LVGenU, such as Wasserstein distance and Tail Capture, but also significantly improved on areas where it was already outperforming the other baseline models. The result of these improvements is plausible substation loads with improved representation of the distribution extremes. LVGenWCS then improves all metrics further to a level where load reconstruction can be carried out with high accuracy.

\begin{figure*}[htb]
    \centering
    \begin{subfigure}[b]{0.48\textwidth}
        \centering
        \includegraphics[width=\textwidth]{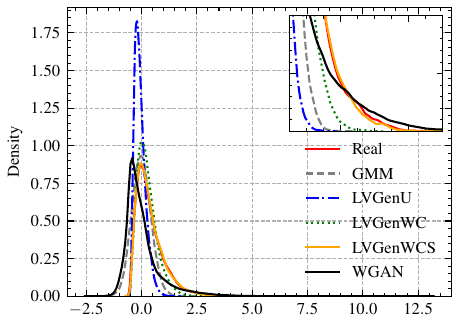}
        \caption{Active Power}
        \label{fig:apdist}
    \end{subfigure}
    \hfill
    \begin{subfigure}[b]{0.48\textwidth}
        \centering
        \includegraphics[width=\textwidth]{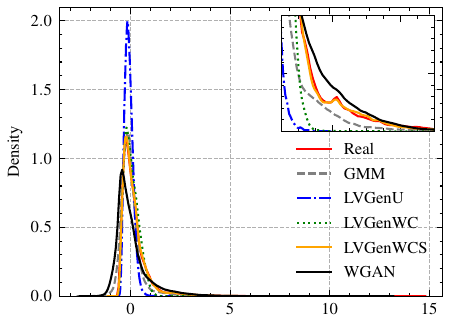}
        \caption{Reactive Power}
        \label{fig:rpdist}
    \end{subfigure}
    \caption{Comparison of the distributions of Active and Reactive Power for different models and the real data. The detail shows how the upper tail behaviour is captured.}
    \label{fig:power_distributions}
\end{figure*}

Figure~\ref{fig:power_distributions} plots the active and reactive power distribution curves of the real dataset and each of the proposed models. Results show a similar trend to what is observed in the Marginal score and Wasserstein Distance; the increase in conditions results in a more accurate capture of the distribution. The GMM provides a good capture of the overall distribution for the active and reactive power; however, it fails to capture both tails of the distribution. This is due to the GMM’s fitted mixture components not adequately capturing low‑probability regions in the unseen test set. The WGAN demonstrates a better overall fit to the distributions compared to the GMM, mainly due to its improved capture of the peak of each distribution. However, it still fails to accurately capture the lower tail. This behaviour reflects the influence of the WGAN’s penalty term, which contributes to its improved Wasserstein Distance and Marginal Score. However, this emphasis results in weaker performance in all other metrics, namely MSE and MMD. LVGenU demonstrates a pessimistic generation rarely diverging from the median samples in order to minimise the MSE loss function. The lower MSE allows the model to achieve better reconstruction of load profiles; however, this comes at the cost of a less accurate distribution capture, highlighted by a higher Marginal score and Wasserstein distance compared to the GMM and WGAN. However, despite this limitation, the lower MMD score achieved by LVGenU compared to the other unconditional baselines indicates that, on a per-substation basis, LVGenU aligns more closely with the real data. The tail-focused MSE metrics shows although models exhibit differences of distributional capture at extremes, their point-wise errors are largely similar. LVGenWC improves upon LVGenU by generating a wider variety of samples and achieving a more accurate capture of the distribution, as demonstrated by an improvement in each of the distance-based metrics and reduced Tail-Focused MSE. This is due to the distribution containing a less significant oversample of the median, and not as severe a miss of the extreme values. However, both issues persist, though not to the same extent as with LVGenU. The distribution generated by LVGenWCS (orange) almost perfectly matches the real distribution (red), specifically at the upper distribution tail, which previous models failed to capture. This is shown by a drastic reduction in Tail-Focused MSE. These results make it difficult to distinguish between the true and generated distributions in the figures, as LVGenWCS closely overlaps the real. The model is also able to capture multi-modal behaviour in the tail of the distribution for the reactive power. This leads to a substantially lower MMD score compared to all other models, representing the most significant improvement in the metric by a wide margin.
\begin{figure*}[!htbp]
    \centering
    \includegraphics[width=1\textwidth]{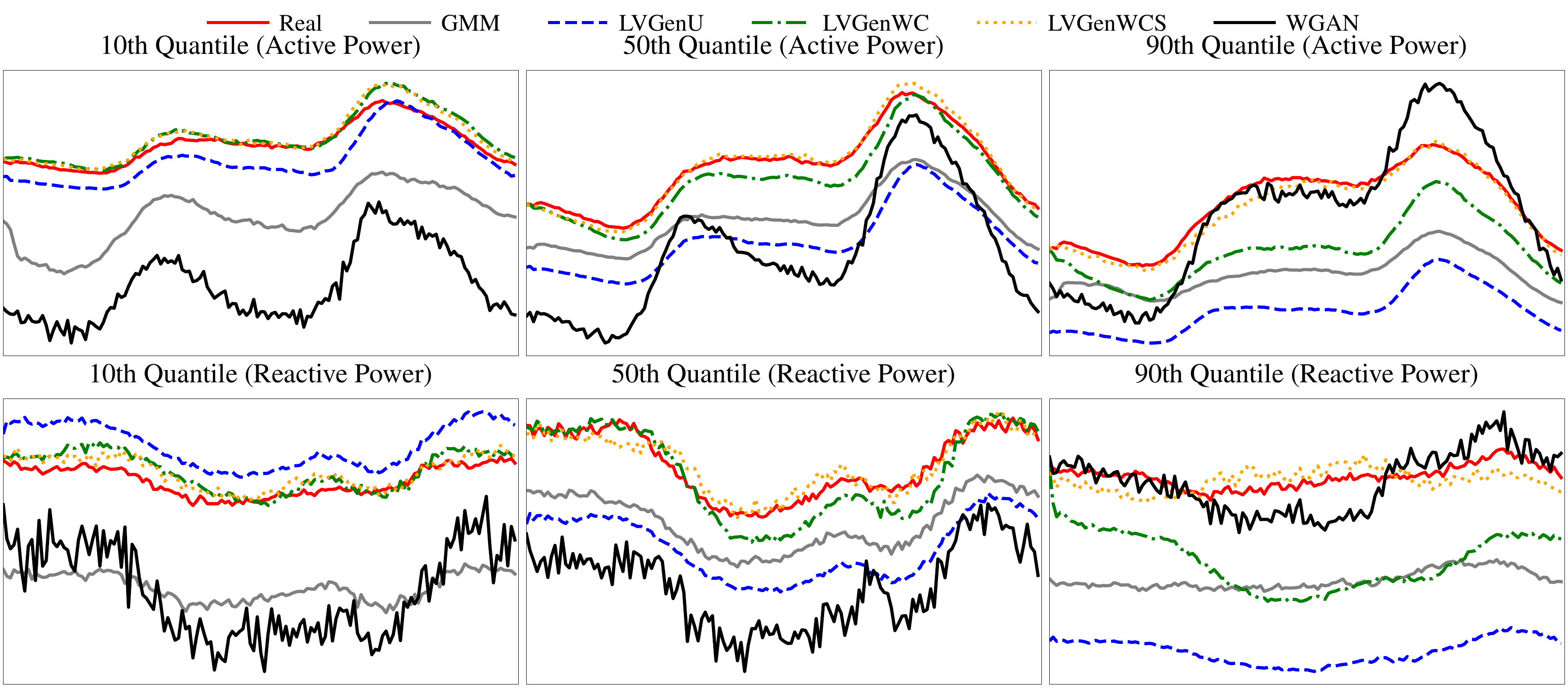}
    \caption{Decile plots for the 10th, 50th, and 90th decile for each point of the daily load. Plots show the real data, and each model's generated samples for active and reactive power. Results show how the general trend of the different distributions is captured by each model.}
    \label{fig:qualtileplots}
\end{figure*}

Figure~\ref{fig:qualtileplots} contains the active and reactive power decile plots for the real dataset and each of the proposed models. The figure plots the typical load profiles throughout various deciles for active and reactive power. Results show the GMM misses the real load profile at every decile for active and reactive power, all to a similar degree; these consistent misses result in a high MSE. WGAN is in the correct range for the 90th decile, but misses the shape of the load profile. In the other deciles, WGAN misses more severely than the GMM, resulting in an even higher MSE due to these extreme missed values. LVGenU is much closer to the real data for active and reactive power at the 10th decile compared to GMM and WGAN. At the 50th decile, similar results to the GMM are obtained. The model completely misses the 90th decile, however still manages to retain a lower MSE when compared to the benchmark models. The generally high MSE for each unconditional model results from their uninformed generation process, as neither model has any indication of the load profile it is attempting to reconstruct (guidance that subsequent models receive through their conditions). Each model's limited ability to capture behaviour in the upper quantiles is directly correlated with the Tail-Focused MSE. LVGenWC provides a more accurate capture of active and reactive power at the 10th and 50th deciles, performing comparably to LVGenWCS in these areas. However, it struggles to capture the 90th decile, a trend similar to what was observed in Figure~\ref{fig:power_distributions}. This is reflected in both MSE metrics, where LVGenWC shows improvement due to its better capture of the 10th and 50th deciles, along with a less severe miss at the 90th decile. LVGenWCS further improves the MSE metrics by achieving a much better capture of the 90th decile for both active and reactive power.

Figure~\ref{fig:acf_comparison} contains the active and reactive power plots for the ACF. Results show all diffusion models are very closely correlated with the real data throughout the lag values for the active power. The GMM loses correlation with the real data for active from lag 25 onwards. For reactive power, the WGAN in particular struggles and loses temporal coherence with the real data. For the other models, including the GMM, the correlation is strong. The MMD metric can provide additional insight into the similarity between the temporal structure of the real and generated data sets. GMM has a lower MMD than WGAN due to WGAN's poor performance on the reactive power. LVGenU further improves upon the MMD performance of GMM, which may be attributed to improvements observed in the ACF plot for active power. The addition of conditional inputs to the diffusion model further improves the MMD, similarly to all other metrics. This highlights the importance of the conditional values for the generation process.

Together, the metrics and visualisations outline the performance of each model and demonstrate the improvement from the baseline GMM and WGAN to a diffusion model in an unconditional perspective. This is through the ability of the diffusion model to capture higher-order moments in the data and better replicate the temporal structure of real data. Then the addition of conditional inputs led to further improvements being observed in the results. The LVGenWC model shows promise, being capable of generating substations from a much greater range of the distribution, accounting for many of the substation loads in the dataset. However, it does not adequately sample from the extremes of the distribution, nor does it consistently recreate load profiles to high accuracy. LVGenWCS is far more suitable as it can generate specific load profiles from individual substations with high quality and accuracy right across the distribution. The results show that the daily statistics were especially required in order to accurately model the extremes of the distribution; substation metadata alone is not enough. While monitoring data is limited targetting specific substations based on geography or other possible factors may provide a potential solution. In subsequent sections, it is demonstrated how the models generated aggregated load profiles perform in a wider power systems context through the application of power systems models. Although informative in terms of model performance, synthesis metrics alone are not sufficient. A model must also produce robust phase angles and voltage estimates across the entire network to serve as a credible replacement for rigorous monitoring.

\section{Power System Analysis Case Study: Urban Scale Load Flow}\label{sec:casestudy}

Results from the Diffusion models, when applied to the synthesis of LV loads, show promising accuracy; to evaluate if generative data could replace real load data, analysis of single loads is insufficient for assessing overall network stability and safety. One important consideration is to ensure that the voltages are within statutory limits to prevent any over or undervoltage conditions. Furthermore, it is important to determine the extent to which the network components, like transformers, cables, or capacitors, may need to be reinforced or upgraded based on the load flow analysis.

To address this, simulations of Medium Voltage (MV) distribution feeders are performed, which are based on representative urban and rural 33kV networks in Great Britain (GB). The feeders are populated with actual LV monitoring data, and load flows are calculated to obtain bus voltage magnitudes and phase angles. Additionally, the `Tao Vanilla' model~\cite{Hong2011} is provided as an additional benchmark model alongside the GMM and WGAN. The Tao Vanilla model is widely used as a benchmark in power system case studies and aligned research to conduct load forecasting and analysis of load flow. The model uses multiple linear regression to predict the load based on a set of load-driving instantaneous criteria, including weather and calendar data:

 \begin{equation}
 \begin{aligned}
 & {\rm E} (Load) =\beta_{0}+\beta_{1}\times Trend+\beta_{2}\times Day\cr &\times Hour+\beta_{3}\times Month + \beta_{4}\times Month\cr &\times TMP+\beta_{5}\times Month\times TMP^{2}+\beta_{6}\cr &\times Month\times TMP^{3}+ \beta_{7}\times Hour\times TMP\cr &+\beta_{8}\times Hour\times TMP^{2}+\beta_{9}\times Hour\cr &\times TMP^{3}
 \end{aligned}
 \end{equation}

Tao Vanilla can express the relationship between load, temperature, and seasonal and diurnal covariates through multilayer linear relationships, making it a good benchmark model for expressing these simple relationships. 

\subsection{United Kingdom Generic Distribution System (UKGDS)}
\begin{figure}[ht]
    \centering
    \includegraphics[width=1\columnwidth]{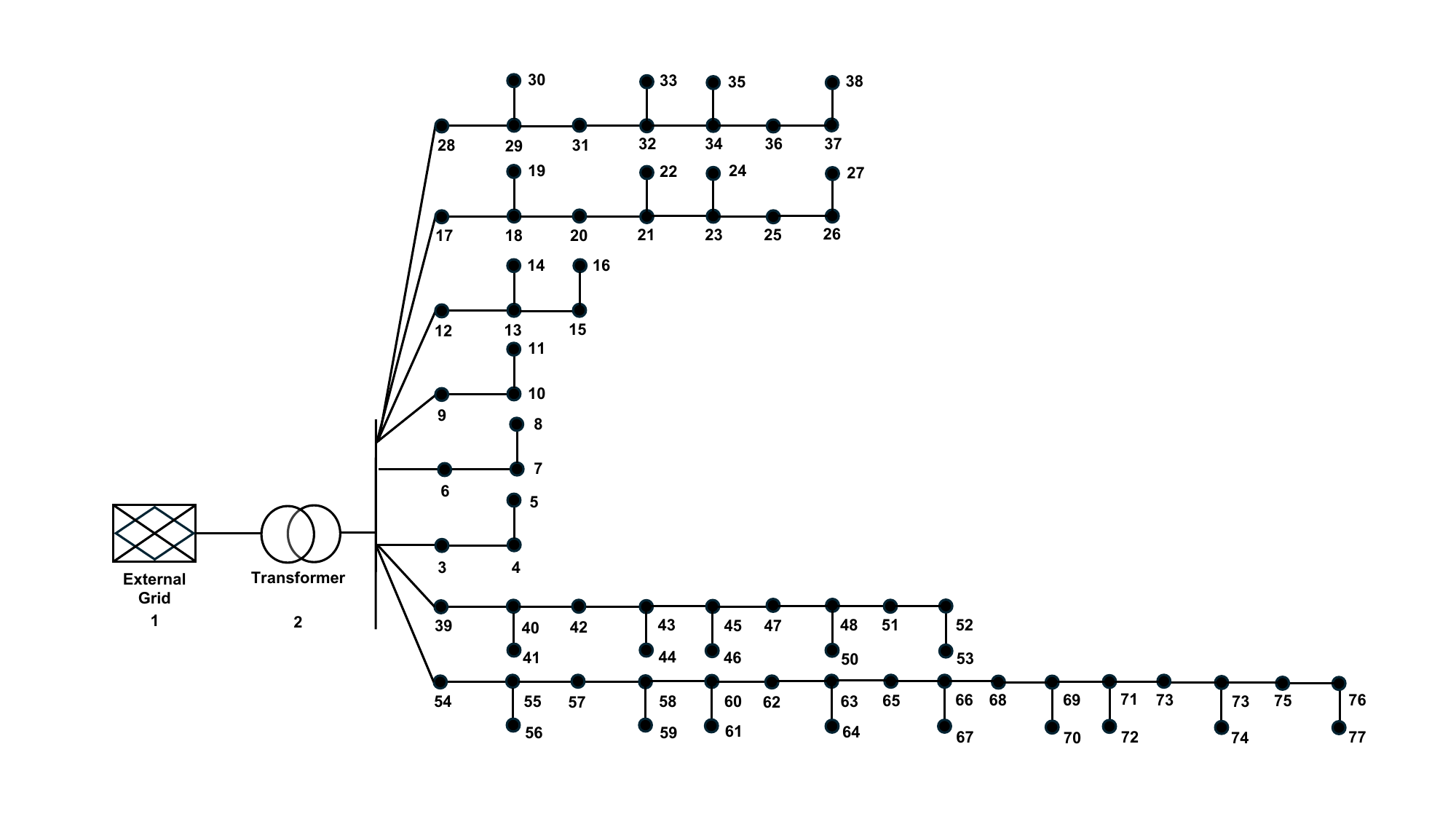}
    \caption{UKGDS Network Test Network used to assess MV impact from LV substation load behaviour.}
    \label{fig:ukgds}
\end{figure}

UKGDS contains several models that can represent the behaviour of the GB distribution networks. In this paper, a UKGDS 77-bus test network is examined. It is simulated for an urban area with a high customer density, containing both 33~kV and 11~kV networks. The structure of the UKGDS network is shown in Figure~\ref{fig:ukgds}.

The network covers an area of approximately $10 \, \text{km}^2$. It consists of one 33~kV substation that acts as the slack bus, with two transformers at the same location, stepping down the voltage from 33~kV to 11~kV. In total, there are 75 loads connected to the remaining 75 11~kV substations distributed throughout the network. The average distance between each bus is about 0.75~km. The network connections are organised into several levels, which enable distant substations to operate at lower voltages. In this network, the designed limit for the voltage magnitude is between 0.97 to 1.03 p.u., and the base capacity is 10~MVA. In this study, the 75 LV substations were analysed using three repeated sets of data consisting of 26 actual LV substations to simulate network performance. This approach aims to maximise the diversity of load flow results across the network. Additionally, only the 26 substations that were active in the network were also used for testing. The results demonstrated similar results to those of the 75 substations, but showed reduced diversity on the low-voltage side, attributed to the lower load within the network.

\subsection{Load Flow Setup and Analysis}
The load flow analysis is used to verify that the power system operates within network limits ($\pm 0.05\,\text{p.u.}$ for voltage magnitude and 10 degrees for phase angle), which are standard values in GB distribution networks. This enables the identification of constraint violations and assessment of system feasibility. The load flow used is based on the Newton-Raphson method and is performed using the Python library \textit{pandapower}~\cite{pandapower}. Given an $n$-bus system with one slack bus, which has a constant 1.0 p.u. voltage magnitude and zero phase angle. The complex power at the node $n$ can be calculated as~\cite{William1967}:
 \begin{equation}\label{William1967}
(P_n + jQ_n) = \overline{E}_n \sum_{m=1}^{n} \overline{Y}_{nm} \overline{E}_m
\end{equation}
where \( P \) and \( Q \) are the active and reactive power injecting into the bus, \( E_n \) is the node-to-datum voltage, \( Y_n \) is the element of the admittance matrix, \textit{j} is a complex number, and \(\overline{E}\) indicates complex quantities. To apply the Newton method, a Jacobian Matrix is used to represent the partial derivatives of the load flow equations, which can be represented as:
\begin{equation}
J = \begin{bmatrix}

\frac{\partial P}{\partial \theta} & \frac{\partial Q}{\partial \theta} \\
\frac{\partial P}{\partial V} & \frac{\partial Q}{\partial V}
\end{bmatrix}
\end{equation}
where \( V \) is the voltage magnitude and \( \theta \) is the phase angle, the voltage magnitude and phase angle can be iteratively updated using the Jacobian matrix by:
\begin{equation}
\begin{bmatrix}
\Delta \theta \\\Delta V\end{bmatrix}
= -J^{-1}\begin{bmatrix}\Delta P \\\Delta Q
\end{bmatrix}
\end{equation}
Therefore, the updated voltage magnitude and phase angle after each iteration is:
\begin{align}
\theta^{(k+1)} &= \theta^{(k)} + \Delta \theta \\
V^{(k+1)} &= V^{(k)} + \Delta V
\end{align}
The iterations continue until the power mismatches and (\(\Delta P\) and \(\Delta Q\)) are within a specified tolerance.
In this analysis, only the voltage magnitude and phase angle are considered, as these are the main parameters of interest to the DNO. Although line losses, thermal margins, and line currents can also be included in power flow studies, the focus here is on assessing whether the synthesised load can represent the network operating conditions. Since the DNO’s main concern is whether these effects are properly captured by the power flow analysis, it is well established that the voltage magnitude and phase angle can be presented as key indicators of network performance \cite{dss}. Other aspects can be explored in future research by generalising the contribution.
\subsection{Load Flow Simulation Results}


\begin{figure}
    \centering
    \includegraphics[width=1\linewidth]{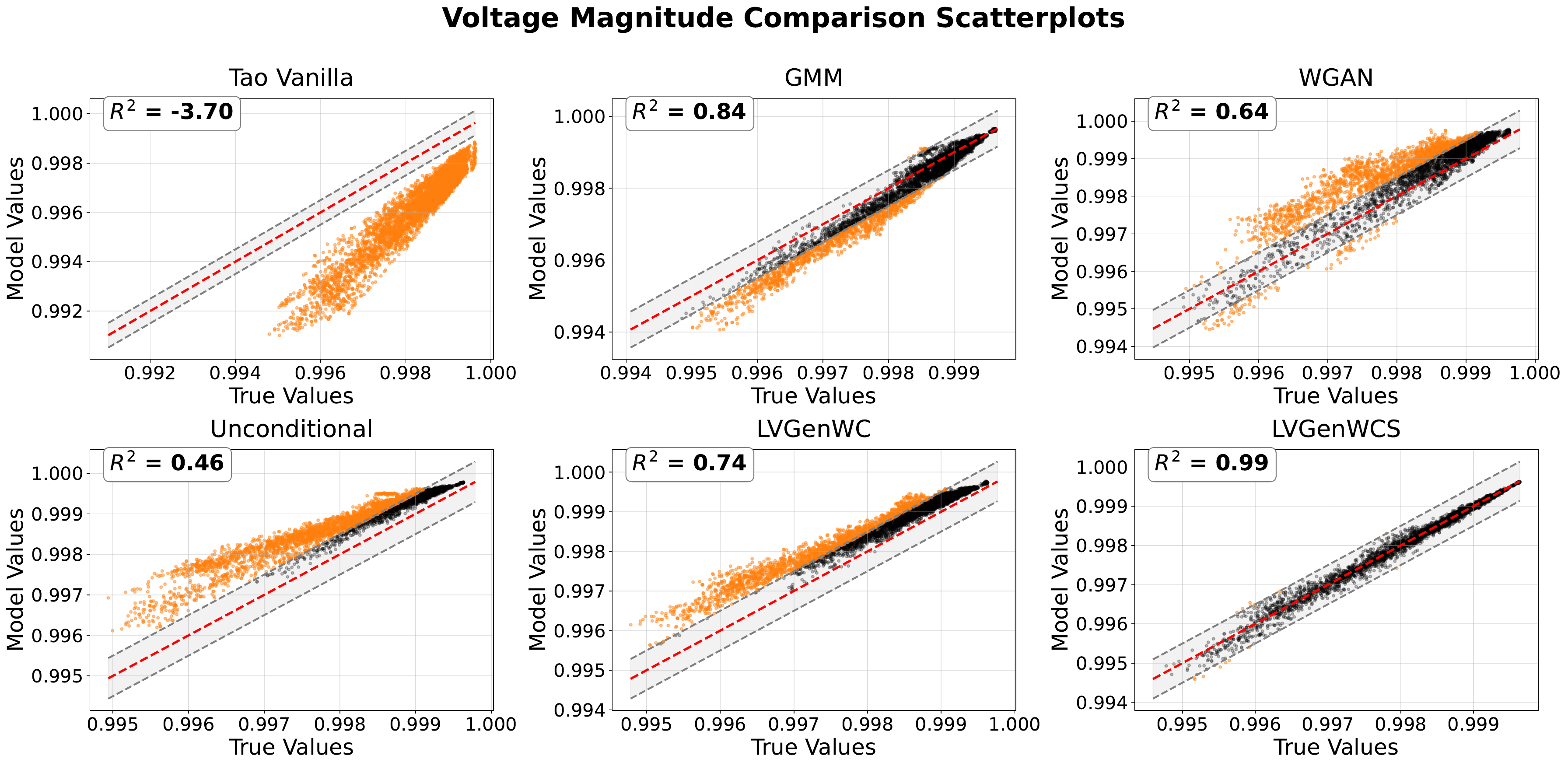}
    \caption{Comparison scatterplot of Voltage Magnitude predictions versus ground truth for each model. Each panel contains acceptance bands representing acceptable deviations from the true values ($\pm1\%$ for voltage magnitude). Points within the acceptable band are plotted in black, while points outside are plotted in orange.}
    \label{fig:voltage_magnitude_error}
\end{figure}


\begin{table}[ht]
\centering
\caption{Comparison error table of Voltage Magnitude predictions versus ground truth for each model.}
\label{tab:Voltage-Magnitude-results}
\resizebox{\textwidth}{!}{%
\begin{tabular}{lcccc}
\toprule
\textbf{Model} & \textbf{MAE (V)} & \textbf{$R^2$} & \textbf{5th Percentile Error (V)} & \textbf{95th Percentile Error (V)} \\
\midrule
Tao Vanilla & 3.13 & -2.48 & 0.536 & 6.8 \\
GMM           & 3.31 & -0.37 & 0.184 & 9.0 \\
WGAN          & 5.28 & -2.96 & 0.633 & 13.5 \\
LVGenU        & 7.03 & -5.47 & 2.605 & 15.4 \\
LVGenWC       & 4.83 & -2.57 & 1.606 & 10.6 \\
LVGenWCS      & \textbf{0.75} & \textbf{0.89} & \textbf{0.030} & \textbf{2.720} \\
\bottomrule
\end{tabular}
}
\end{table}
Figure~\ref{fig:voltage_magnitude_error} and Table~\ref{tab:Voltage-Magnitude-results} present a comparison of the voltage magnitude results obtained from the benchmark and proposed diffusion models. To ensure statistical robustness and account for the stochastic nature of the generative process, all LVGen models were evaluated over 10 independent runs; the results reported herein represent the average values across these runs. In Figure~\ref{fig:voltage_magnitude_error}, the red reference line represents perfect agreement between the true and predicted values. The black dashed lines define an acceptance band corresponding to $\pm1\%$ of the voltage magnitude, in accordance with statutory limits. Points within the band are plotted in black, while points outside the band are highlighted in orange.

The models generally perform well in estimating higher voltage magnitudes when the system load is at a minimum, but tend to be less accurate at lower voltage magnitudes when the load is near its peak. In general, most methods overestimate the voltage magnitude, which could result in inappropriate decisions by system operators. In contrast, the LVGenWCS method consistently captures the trend of the voltage magnitude accurately, regardless of whether the values are high or low. 

The results in Table~\ref{tab:Voltage-Magnitude-results} compare the six methods, which show that LVGenWCS demonstrates a MAE that is five times lower (than the next best model). In addition to MAE, the 5th and 95th percentile errors are also reported. The range between the 5th and 95th percentiles provides insight into the distribution of the majority of errors, effectively excluding the most extreme 5\% at either end. Such insights are especially valuable for power system operators, whose primary concern is the reliable and consistent performance of predictive models under typical operating conditions. The LVGenWCS method also exhibits substantially lower errors within the 5th to 95th percentile range, indicating greater accuracy and reliability across the full spectrum of voltage magnitudes. Moreover, the results suggest that, in the context of a constrained power system where the system operator aims to maintain the voltage magnitude within ±0.05 p.u., only the LVGenWCS method possesses the capability to provide load predictions that remain within this limit. Consequently, it is the only method that can reliably support the system operator in making appropriate operational decisions.

\begin{figure}[!htbp]
    \centering
    \includegraphics[width=0.7\columnwidth]{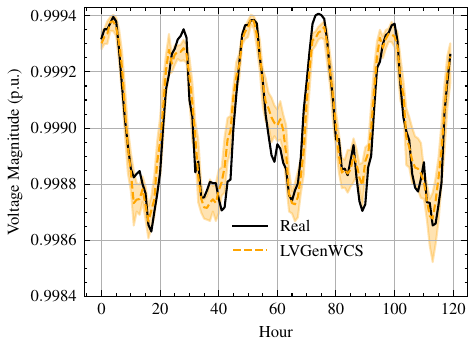}
    \caption{Single LV bus time series of hourly average voltage magnitude demonstrating temporal fidelity between real and LVGenWCS data with min and max thresholds. Variability in model output over multiple stochastic generations is highlighted using min-max error bars.}
    \label{fig:Voltage Magnitude}
\end{figure}

Furthermore, Figure~\ref{fig:Voltage Magnitude} compares the LVGenWCS with the real data for a single bus test case. The LVGenWCS model results are shown with confidence intervals also reported based on the model's upper and lower outputs at each timestep. The findings demonstrate that the LVGenWCS model provides reliable results that replicate the dynamic behaviour of the network. The LVGenWCS model closely aligns with the real load data, accurately capturing voltage patterns with reduced deviations in amplitude. Although some of the troughs are overestimated in certain model results, the average of the results is sufficiently accurate, with an acceptable error margin. This indicates that the LVGenWCS model offers reliable accuracy for the network and serves as a good fit for predicting voltage magnitudes.
\begin{figure}
    \centering
    \includegraphics[width=1\linewidth]{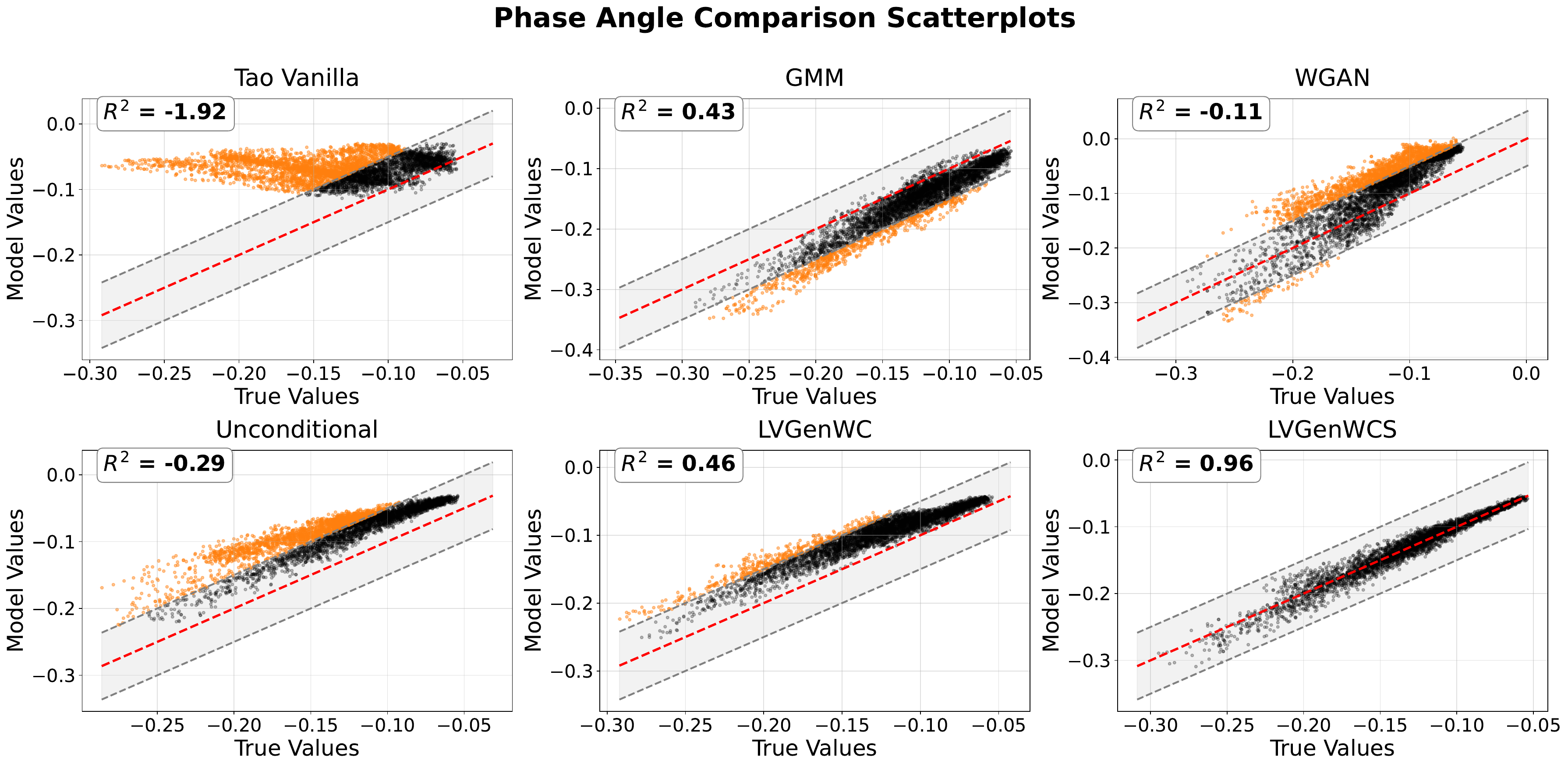}
    \caption{Comparison scatterplot of Phase Angle predictions versus ground truth for each model. Each panel contains acceptance bands representing acceptable deviations from the true values ($\pm5\%$ for phase angle). Points within the acceptable band are plotted in black, while points outside are plotted in orange. }
    \label{fig:phase_angle_predictions}
\end{figure}


\begin{table}[ht]
\centering
\caption{Comparison error table of Phase Angle predictions versus ground truth for each model.}
\label{tab:phase-angle-errors}
\resizebox{\textwidth}{!}{%
\begin{tabular}{lcccc}
\toprule
\textbf{Model} & \textbf{MAE (deg)} & \textbf{$R^2$} & \textbf{5th Percentile Error (deg)} & \textbf{95th Percentile Error (deg)} \\
\midrule
Tao (Vanilla) & 0.056 & -2.80 & 0.02 & 0.09 \\
GMM           & 0.028 &  0.19 & 0.005 & 0.06 \\
WGAN          & 0.041 & -0.89 & 0.007 & 0.08 \\
LVGenU        & 0.046 & -1.25 & 0.023 & 0.08 \\
LVGenWC       & 0.029 &  0.01 & 0.01 & 0.05 \\
LVGenWCS      & \textbf{0.007} & \textbf{0.93} & \textbf{0.0004} & \textbf{0.02} \\
\bottomrule
\end{tabular}%
}
\end{table}

Figure~\ref{fig:phase_angle_predictions} and Table~\ref{tab:phase-angle-errors} present a comparison of the Phase Angle results obtained from the benchmark and proposed diffusion models. The results show the Tao Vanilla model performs poorly as it is not designed to predict the reactive power. The results follow similar trends to those observed with voltage magnitude, where models excluding LVGenWCS provide acceptable results for higher values, but display significant bias errors for smaller values. The LVGenWCS model is consistently able to capture both higher and lower values, although the error in smaller values is greater than what was observed with the voltage magnitude results. This indicates that reactive power is considerably more difficult to predict.

The results in Table~\ref{tab:phase-angle-errors} also show that the LVGenWCS method outperforms the other methods, again demonstrating an MAE that is roughly five times lower than the next best model, and ten times lower for the 5th percentile. The reactive power values here are considerably low due to the data only consisting of resident loads. The results indicate that only the LVGenWCS method shows potential for wider application in reactive power prediction for networks containing commercial or industrial loads, thereby providing system operators with more reliable day-ahead predictions.

\begin{figure}[ht]
    \centering
    \includegraphics[width=0.7\columnwidth]{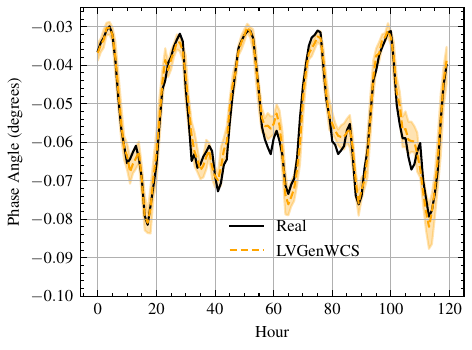}
    \caption{Single LV bus time series of hourly average phase angle demonstrating temporal fidelity between real and LVGenWCS data with min and max thresholds. Variability in model output over multiple stochastic generations is highlighted using min-max error bars.}
    \label{fig:Phase Angle}
\end{figure}

Figure~\ref{fig:Phase Angle} compares the phase angles over time for a single bus between real values and the LVGenWCS model. The results again show similar trends to the voltage magnitude analysis. The LVGenWCS model results closely follow the trend of the real data, effectively capturing both amplitude and frequency of the oscillations. The LVGenWCS model exhibits good deviations from the real data, particularly at the peaks and troughs.

Overall, improvements in model performance with the introduction of initial conditioning in the LVGenWC model do not extend to the power flow case study. This shortcoming does not translate to the network being generated being unrealistic or not plausible, but rather that it does not resemble the true network in this example. In order to accurately model the network in the load flow analysis, the daily load statistics were required. The strong correspondence between the real data and LVGenWCS model outputs for both voltage magnitude and phase angle highlights the effectiveness of the LVGenWCS model in replicating complex voltage characteristics based on the load profile. This high level of accuracy suggests that the model can be reliably used for both prediction and analysis in distribution power systems. It is expected that access to a wider range of reliable metadata in the future, and its subsequent incorperation to the models can limit the requirement for monitoring to perform accurate power flow analysis.  
\clearpage
\section{Conclusion and Future Work}\label{sec:conclusion}

Transitioning towards Distribution System Operation in legacy power systems requires an understanding of load interactions at the sub-regional level to avoid thermal and voltage constraint violations. Extensive monitoring would ultimately inform mitigating strategies to a high standard, but it is prohibitively expensive to deploy and maintain comprehensively across distribution networks owing to the sheer volume of assets. While load behaviour synthesis offers a convenient alternative, demand characteristics at low voltage substations must be realistically diverse so that higher voltage level behaviour is accurately represented: lack of diversity results in voltage collapses and thermal exceedances, excessive diversity will excessively smooth out any extremes. 

To address the problem of realistic distribution load profile synthesis, this work has contributed various Generative Diffusion Models capable of reconstructing LV substation load profiles based on a range of conditional cues. Realism was demonstrated on an individual level through the validation of temporal and statistical characteristics, while coherence in the wider power system context was demonstrated through the propagation of synthesised LV loads through a representative MV network. 

The key outcomes of this study are a fleet of models that contrast unconditional synthesis with progressively conditioned generation to support a range of practical use cases. Results showed that while an unconditional model was capable of generating realistic load profiles, it was not able to synthesise specific load behaviours, making it of limited use from both a scenario generation and reconstruction perspective.  The introduction of conditional metadata in LVGenWC, such as calendar information or weather variables that are often readily available, greatly improved synthesis quality and enabled the model to realistically synthesise a far greater range of behaviours representing a significant number of substations. These models provide a synthesis of substation base load profiles that is suitable for downstream power system studies and serve as a prerequisite for benchmarking the impacts of LCT penetration or changes in customer numbers. However, LVGenWC struggled at distribution extremes, causing it to perform poorly in the load flow analysis. While LVGenWC can generate a network of plausible loads, it struggles to recreate real networks due to mismodelling of individual substations, indicating that conditioning on lightweight non-substation specific metadata alone is insufficient to fully capture behaviour at the distribution extremes. The introduction of daily statistics (daily min, mean, and max of load profiles) improved the model's capture at the distribution extremes, and load flow analysis demonstrated a similar impact on the network to that observed using real meter data. Although such data may not always be available, these models are capable of generating highly accurate networks which could be used for various other tasks within the power system, such as stress testing.

There are a number of areas of limitation which future work aims to address. The results show that although a conditional diffusion model driven by substation metadata and weather is capable of generating plausible univariate substation load profiles, the associated performance gains do not consistently translate into realistic network-level behaviour in the power flow case study. In particular, when substations are generated independently and without synchronisation, additional statistical conditioning is required to reproduce physically meaningful aggregate behaviour. This observation motivates future work aimed at improving the capture of inter-substation co-behaviour and enabling the modelling of LV networks without reliance on extensive monitoring data.

The present study focuses on modelling LV base load and does not explicitly capture the distinct dynamics associated with high-penetration LCTs. This choice reflects the current lack of reliable, large-scale LCT datasets suitable for conditioning deep generative models. As such datasets become available, the proposed framework can be retrained with additional conditional priors to explicitly represent LCT-driven demand dynamics and support deep-decarbonisation scenario analysis. Furthermore, the models are trained and tested on a single DNO dataset, and their evaluation must be extended to datasets from other DNOs and geographical contexts to ensure generalisability in other regions, network topologies and customer mixes.

Moreover, the evaluation primarily assesses model performance under historical operating conditions. Identifying and characterising structural breaks or regime changes in LV load behaviour is a research challenge in its own right and would require a dedicated methodological contribution. While testing on unseen substations provides a partial mitigation of domain shift, explicitly evaluating model robustness under structural change remains an open challenge and a key direction for future work.

Finally, the proposed diffusion-based framework is inherently data-driven and does not explicitly embed physical or first-principles load models. While power-flow-based evaluation demonstrates that the independently synthesised profiles produce physically plausible network behaviour, the internal representations learned by the model remain difficult to interpret. Improving model transparency and explainability, particularly in relation to physically meaningful power-system quantities, is therefore an important direction for future research.

\section*{Acknowledgments}
Results were obtained using ARCHIE-WeSt High Performance Computer (www.archie-west.ac.uk).

\color{black}

\bibliographystyle{unsrt}
\bibliography{references}

\appendix
\section{WGAN Training}\label{sec:app}


This section will cover the training of the Wasserstein GAN (WGAN) model used as a benchmark in the study. The WGAN model used was implemented by Hao et al.~\cite{DBLP:journals/corr/abs-2111-01207}. Like with SSSDS4, the model was used with existing hyperparameters with one minor modification. A non-autoregressive generator was added to focus the model on synthesis rather than forecasting. 

\begin{figure}[htbp]
    \centering
    \includegraphics[width=0.8\linewidth]{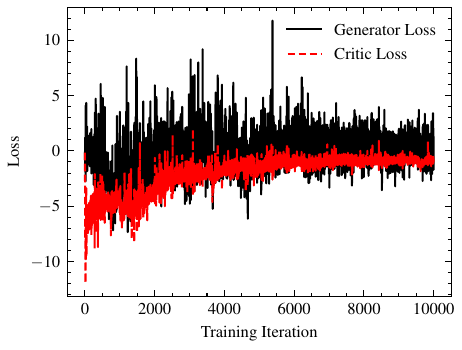}
    \caption{Loss terms for the Generator and Critic for each training step.}
    \label{fig:GenLoss}
\end{figure}

\begin{figure}[htbp]
    \centering
    \includegraphics[width=0.8\linewidth]{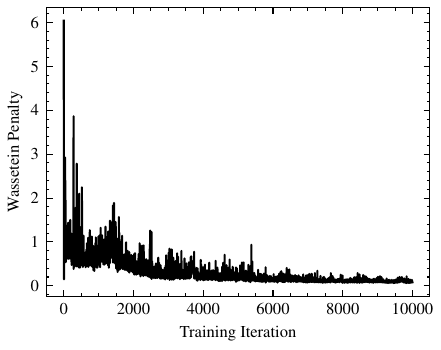}
    \caption{Wasserstein Loss penalty term for each training step.}
    \label{fig:wassloss}
\end{figure}

Generally, training of a GAN can be challenging to know when to stop the model based on its multi-objective functions ~\cite{gao2025evolution}. Specifically in the WGAN, where the extra loss penalty can act as another optimiser. For training, the model was tested to 100 \& 200 epochs and also stopped when visual inspection of the loss terms looked favourable. The deviations in results between each other were minimal, but the earlier-stopped model provided slightly improved results. Figures~\ref{fig:wassloss} \& \ref{fig:GenLoss} show the loss terms during training. Given the unconditional nature of the model the training converged quite quickly once the Wasserstein penalty had stabilised; there was little extra information for the model to learn. Out of the scenarios described above, the best results were obtained when stopping the model around 2000 iterations, when the generator's performance was better. However, it must be noted that the difference in final results and metrics is minimal. 

\section{SSSDS4 Masking Mechanism}\label{sec:appmask}
This section will provide a detailed explanation of the role of the binary mask in the implementation of the SSSDS4 model to permit conditional synthesis. SSSDS4, in its existing state, utilises a masking mechanism to conceal inputs from the model during training. The configuration of the mask is what specifies the specific problem the model is trying to solve. The mask mirrors the same shape as the input data, where every channel and timestep has an associated mask value. When the mask value is 1, the original input is provided to the model as a condition; when it is 0, the input is replaced with noise, and the model must learn to denoise it. 

\begin{figure}
    \centering
    \includegraphics[width=1\linewidth]{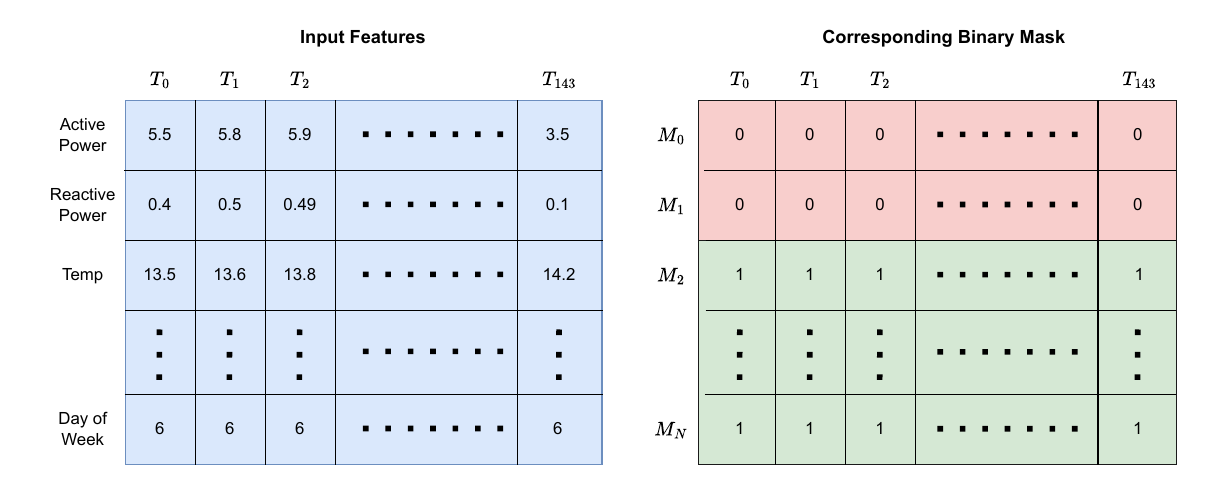}
    \caption{Diagram detailing a sample being passed to the model (left) and its corresponding binary mask used for conditioning (right)}
    \label{fig:condmask}
\end{figure}

Pre-existing masks exist for tasks such as imputation, where values in the mask may randomly be set to 0 across the time dimension for each sample. Alternatively, a continuous block at the end of the timesteps dimension could be set to 0 to perform forecasting. In this study, entire channels are set to either synthesis targets or conditional values, where all timesteps and channels are assigned a mask value of 1 if they are a condition, or 0 if they are a synthesis target. This means the active and reactive power channels have a mask of 0, and all remaining channels are conditional with masking values of 1. Figure~\ref{fig:condmask} shows an example of this configuration of the mask for a single sample being passed to the model. The resulting model will learn a transformation from the conditional inputs and the signal noise vector to a novel active and reactive power load profile. 

\section{SSSDS4 Model Configuration}\label{sec:appconfig}
This section defines the model configuration used in the training of diffusion models in this study. The configuration aligns with parameters used in the original research paper~\cite{alcaraz2023diffusionbased}, which conducted a hyperparameter search including evaluation across multiple datasets. The key training parameters are shown in Table~\ref{tab:modelconfig}. Note that training time will vary depending on GPU capability. The computational complexity and model storage requirements of the implemented models are then summarised separately in Table~\ref{tab:modelcomplexity}.

\begin{table}[H]
\centering
\begin{tabular}{ll}
\hline
\textbf{Parameter} & \textbf{Specification} \\
\hline
Optimizer & AdamW \\
Learning rate & $2 \times 10^{-4}$ \\
Batch size & 100 \\
Maximum epochs & 200 \\
Stopping criterion & Validation loss plateau \\
Validation loss frequency & 10 epochs \\
Patience & 2 \\
Loss function & MSE \\
Number of diffusion steps ($T$) & 200 \\
$\beta_0$ & 0.0001 \\
$\beta_T$ & 0.02 \\
GPU & NVIDIA A40 \\
\hline
\end{tabular}
\caption{Training configuration for implementation of the SSSDS4 model}
\label{tab:modelconfig}
\end{table}
\begin{table}[H]
\centering
\begin{tabular}{lcc}
\hline
\textbf{Model} & \textbf{Single Inference GOPS} & \textbf{Total Trainable Parameters} \\
\hline
LVGenU   & 2.766 & 47.923 M \\
LVGenWC  & 2.825 & 48.334 M \\
LVGenWCS & 2.857 & 48.559 M \\
\hline
\end{tabular}
\caption{Model storage and computational complexity. Inference throughput is reported in giga-operations per second (GOPS), and total trainable parameters are reported in millions (M).}
\label{tab:modelcomplexity}
\end{table}

\clearpage
\section{Addional Results}\label{sec:AdditionalResults}
This section contains additional insight into error metrics, recording the metrics separately for both active and reactive power. The results are shown in Table~\ref{tab:errors_ap_rp}.
\begin{table}[H]
    \centering
    \resizebox{\textwidth}{!}{%
        \begin{tabular}{llccccc}
            \toprule
            \textbf{Power Type} & \textbf{Metric} 
            & \textbf{GMM} 
            & \textbf{WGAN} 
            & \textbf{LVGenU} 
            & \textbf{LVGenWC} 
            & \textbf{LVGenWCS} \\
            \midrule

            \multirow{6}{*}{\textbf{Active Power (AP)}} 
            & \textbf{MSE}                    & 0.70 & 1.30 & 0.63 & 0.17 & 0.05 \\
            & \textbf{MMD}                    & 0.32 & 0.46 & 0.29 & 0.07 & 0.02 \\
            & \textbf{Marginal Score}         & 0.38 & 0.11 & 0.11 & 0.12 & 0.02 \\
            & \textbf{MiVo}                   & 8.0 & 11.1 & 7.3 & 3.7 & 2.0 \\
            & \textbf{Wasserstein Distance}   & 0.39 & 0.41 & 0.45 & 0.15 & 0.01 \\
            & \textbf{Tail-Focused MSE}       & 5.0 & 5.1 & 5.2 & 1.7 & 0.3 \\
            \midrule

            \multirow{6}{*}{\textbf{Reactive Power (RP)}} 
            & \textbf{MSE}                    & 1.15 & 1.34 & 0.74 & 0.41 & 0.08 \\
            & \textbf{MMD}                    & 0.30 & 0.37 & 0.22 & 0.08 & 0.03 \\
            & \textbf{Marginal Score}         & 0.10 & 0.07 & 0.21 & 0.12 & 0.10 \\
            & \textbf{MiVo}                   & 8.1 & 9.0 & 6.4 & 3.8 & 2.0 \\
            & \textbf{Wasserstein Distance}   & 0.19 & 0.15 & 0.30 & 0.15 & 0.02 \\
            & \textbf{Tail-Focused MSE}       & 10.5 & 10.1 & 10.2 & 4.3 & 0.6 \\
            \bottomrule
        \end{tabular}%
    }
    \caption{Alternative error metric table containing separate performance for active and reactive power.}
    \label{tab:errors_ap_rp}
\end{table}

\end{document}